\documentclass[twocolumn,showpacs,aps,prl,amsmath,amssymb,endfloat]{revtex4-1}
\usepackage{graphicx}             

\begin{document}

\title{Topological Conditions of Scale-free Networks for Cooperation to Evolve}

\author{Dong-Ping Yang}
\email{dpyang@xmu.edu.cn}
\author{Hai Lin}
\author{Chen-Xu Wu}
\author{J W Shuai}
\email{jianweishuai@xmu.edu.cn}
\affiliation{Department of Physics and Institute of Theoretical Physics and Astrophysics, Xiamen University, Xiamen 361005, P.R.China}

\date{\today}

\begin{abstract}
Evolutionary game theory is employed to study topological conditions of scale-free networks for the evolution of cooperation. We show that Apollonian Networks (ANs) are perfect scale-free networks, on which cooperation can spread to all individuals, even though there are initially only 3 or 4 hubs occupied by cooperators and all the others by defectors. Local topological features such as degree, clustering coefficient, gradient as well as topology potential are adopted to analyze the advantages of ANs in cooperation enhancement. Furthermore, a degree-skeleton underlying ANs is uncovered for understanding the cooperation diffusion. Constructing this kind degree-skeleton for random scale-free networks promotes cooperation level close to that of Barab\'asi-Albert networks, which gives deeper insights into the origin of the latter on organization and further promotion of cooperation.
\end{abstract}


\maketitle
Cooperation phenomena are ubiquitous and essential in natural and human systems \cite{Axelrod03271981,*Fehr2003}. Since selfish actions provide a short-term higher benefit, evolutionary game theory \cite{Smith1982,*Nowak2006a} has been employed widely to understand how and why cooperation emerges and survives. The Prisoner's Dilemma (PD) game captures the essence of this problem. In PD game, each individual adopts cooperation or defection. Mutual cooperation or mutual defection provides them both $R$ or $P$, respectively, while a defector can gain largest benefit $T$ from a cooperator, which in turn receives $S$. From the order $T>R>P\geq S$, it is obviously better to defect, regardless of the opponent's strategy. In the well-mixed population, a strategy receiving higher payoffs spreads within a population and the proportion of cooperators vanishes asymptotically.

The interaction structure in real societies are more naturally to be described by complex networks and various networks have been reported to provide an asymptotic survival of cooperation \cite{Santos2006pnas,*ohtsuki06,*Szabo200797}. Notably, cooperation is enhanced significantly in scale-free networks \cite{Santos2005,Gomez2007}. Especially on the so-called Barab\'asi-Albert (BA) networks \cite{Barabasi1999}, the strong correlation between individuals renders cooperation as the dominating trait. While progress has been made in understanding general properties of social networks for enhancing cooperation \cite{Assenza2008,Floria2009,Devlin2009a,Pusch2008,*Devlin2009b,*Chen2007379}, the question why BA networks possess such a great potential of cooperation enhancement remains poorly understood. Furthermore, is there any better network for cooperation to evolve and what are the significant topological conditions in scale-free networks?

In this Letter, we start from discussing perfect networks for cooperation to evolve: Apollonian Networks (ANs) \cite{Andrade2005}. We discover that cooperation can spread over the whole population on ANs throughout the entire Pareto-efficient parameter range ($T+S<2R$), even though initially only 3 or 4 most-linked nodes (hubs) are occupied by cooperators and all the others by defectors. Then, a hidden degree-skeleton is introduced and demonstrated to be essential for cooperation spreading and further enhancement on ANs as well as BA networks. Constructing a degree-skeleton for random scale-free networks \cite{Molloy1995} promotes the cooperation level close to that of BA networks. Our findings deepen the insights into the origin of BA networks on cooperation organization and enhancement.


ANs are constructed by adding a new node inside one of the existing triangles orderly, starting from a triangle. They are grown generation by generation and in each generation new nodes are just inserted into each triangle created in the last generation. The new node is attached to the three vertices of this triangle, which create three new triangles for next generation. We perform simulations of learning/replicator dynamics on ANs with size $N=10^4$. Note that ANs with $10^4$ nodes are not complete for 10 generations \cite{Andrade2005}. Whether ANs are complete or not makes no difference.

Following common practice \cite{Nowak1992}, parameters of the PD game are chosen as $R=1$, $P=S=0$, and $T=b>1$, where $b$ represents the advantage of defectors over cooperators and $b \leq 2$ means Pareto-efficient. Once ANs are constructed, the dynamics is carried out. Initially, cooperators and defectors are distributed randomly in the network with equal probability unless noted otherwise. At each round, each agent $i$ plays once with all its neighbors and accumulates payoffs stored in $\pi_{i}$. Then, all individuals update their strategies synchronously as well as asynchronously by the following rules: Each individual $i$ selects at random a neighbor $j$ and only if $\pi_{j}>\pi_{i}$, the player $i$ adopts the strategy of its neighbor $j$ with probability $\Pi_{i\rightarrow j} = (\pi_{j}-\pi_{i})/\max \{k_{i},k_{j}\} b$ for next round robin.

For simulations, let the system evolve for a transient time of $5\times10^{4}$ rounds and then several time windows of $10^{3}$ rounds for checking whether the evolutionary dynamics reaches a stationary regime. In each time window, we record down the time evolution of cooperators proportion $\rho_{C}(t)$. If the relative fluctuation $F=\sqrt{\overline{\rho_{C}(t)^{2}}-\overline{\rho_{C}(t)}^{2}}/\overline{\rho_{C}(t)}$ is not larger than 0.008 and the slope of $\rho_{C}(t)$ is smaller that $10^{-3}$, we stop the simulation and employ the average cooperation $\rho_{C}$ obtained in the last time window to be the asymptotic cooperation level. All results reported below are averaged over $10^3$ different realizations of networks and initial conditions.


\begin{figure}
  \includegraphics[width=2.2in]{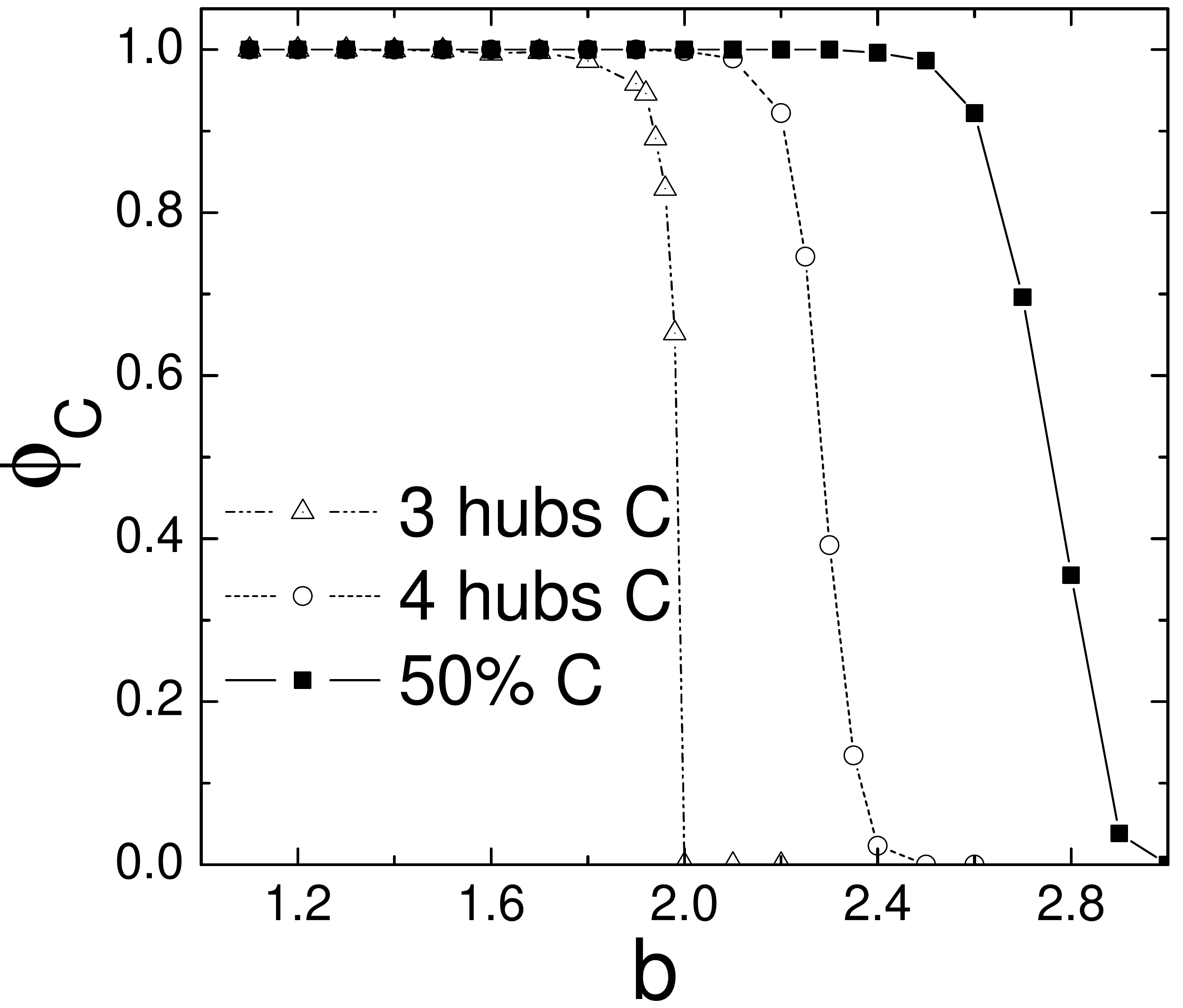}\\
\caption{(a) The probability $\phi_{C}$ that the final state is  $\rho_{C}=1$ as a function of $b$ on ANs for three different initial states: only first 3 hubs, only first 4 hubs, and randomly half of nodes, occupied by cooperators. } \label{AN}
\end{figure}

On ANs the population evolves definitely into two final states: $\rho_{C}=1$ or $\rho_{C}=0$. Fig.~\ref{AN} shows the dependency of $\phi_{C}$ on the defector's temptation $b$, where $\phi_{C}$ denotes the probability that the population reaches the final state of all cooperators ($\rho_{C}=1$). All cooperation can be reached with certainty on ANs up to $b=2.4$ and then $\phi_{C}$ decreases to 0 rapidly when $b$ increases up to 3, in the case that half of the population behave as cooperators initially \footnote{All cooperation can also definitely be reached throughout the entire parameter range of snowdrift game on ANs.}. Surprisingly, just first 3 or 4 hubs initially occupied by cooperators and all other nodes by defectors can induce all cooperation for $b<2$. These results imply that the hubs are extremely stable for cooperation with considerable powerful influences on their neighbors and even the whole population. The outstanding potential for cooperation to evolve on ANs drives us to discuss their underlying topological properties in the following.

First of all, ANs are scale-free networks with low degree exponent: $P(k)\sim k^{-\gamma}$ with $\gamma=1.59$ \cite{Andrade2005}. By the nature of PD game and updating rule, the most prominent mechanism of scale-free networks for promoting cooperation lies in the stability of cooperation on hubs \cite{Santos07012006}. A defector on the hub often induces most of his neighbors to be defectors and reduces the payoff of himself in return. Thus it can be easily invaded by cooperating neighbors with higher payoffs. On the contrary, cooperators on hubs strengthen their stability via assimilating their neighbors to be cooperators and get higher payoffs. Especially on networks with strong age-correlations, interconnections of hubs contribute furthermore the dominance of cooperation for $b<2$ \cite{Santos2005}, such as BA networks ($\gamma=3.0$) generated via growth and preferential attachment. Notice that growth and preferential attachment can be also obtained via local rules of attachment, such as the minimal model \cite{Dorogovtsev2001} and ANs. Furthermore, due to the lower degree exponent, the larger hubs in ANs will strengthen this effect, which results in further enhancement of cooperation. But we will show below that it just partly gives rise to our results in Fig.~\ref{AN}.

Secondly, ANs are highly clustered with clustering coefficient $C_{cl}=0.83$ and the correlation $C_{cl}(k) \sim k^{-1}$ \cite{Andrade2005}. The clustering coefficient \cite{Newman2003} describes the connection probability of two neighbors. Enhancement of cooperation in highly clustered scale-free networks has also been discussed in detail in Ref.\cite{Assenza2008}. For networks with $C_{cl}$ close to 0, the conversion of cooperation into defection is explained as a progressive invasion of the degree classes by defectors: The larger the value of $b$, the more degree hierarchies defectors have invaded \cite{Gomez2008296}. High degree nodes are more resistant to defection and there is a well-defined minimum preference of cooperation for intermediate degree classes. For highly clustered scale-free networks, in contrast, the density of triangles around hubs enhances the fixation of cooperation in low degree nodes when $b$ is low, but homogenizes the invasion process of degree classes by defectors and then decreases the survival of low densities of cooperators at large $b$. For this reason, cooperation can be fixated over the whole population at low $b$ and go to extinction when $b$ is high. The following discussion will be persuasive on this point.

\begin{figure}
  \includegraphics[width=3.2in]{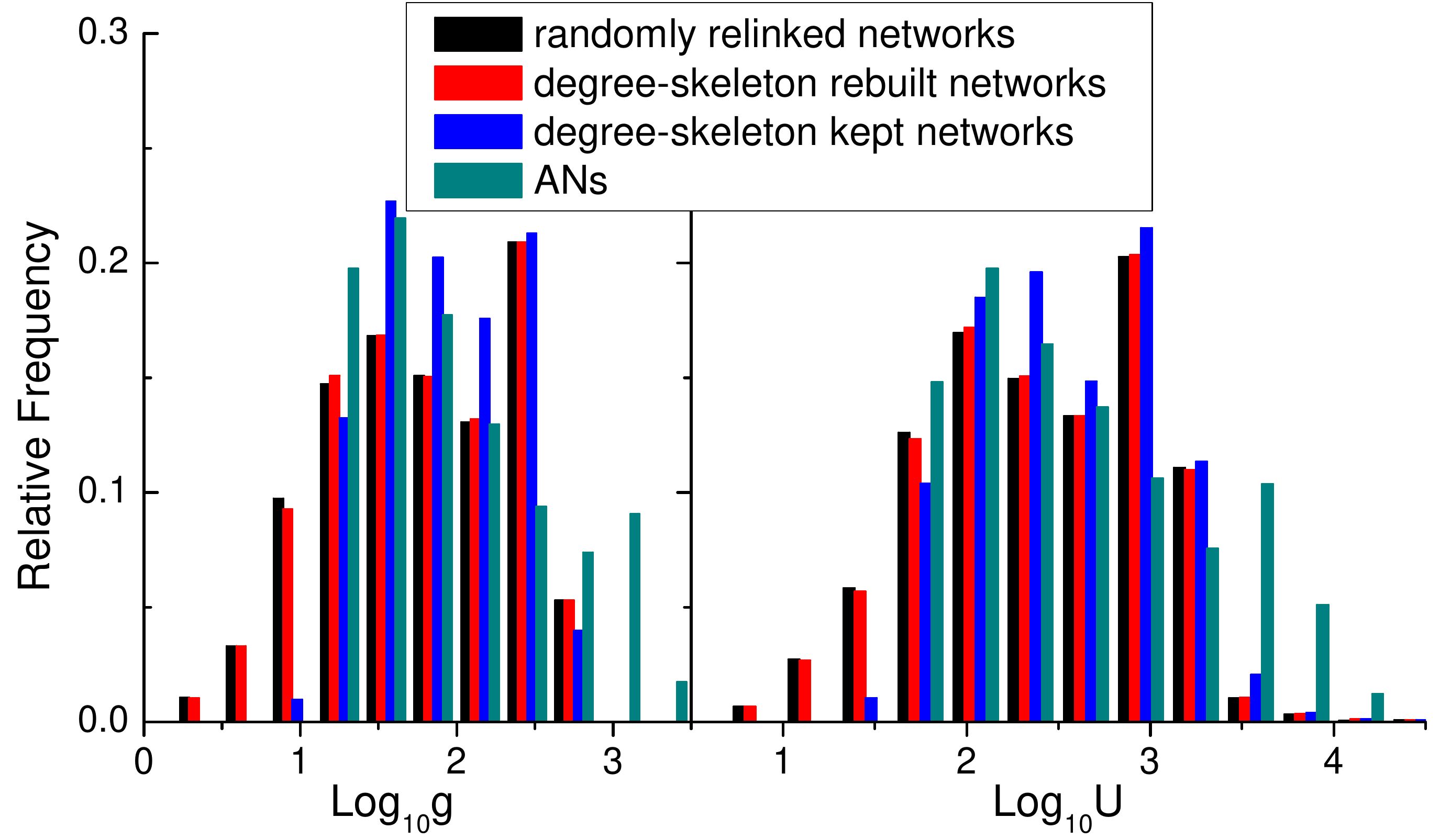}
  \caption{(Color online) Distributions of gradient $g$ and topology potential $U$ for four different networks with identical degree distribution. The first three are derived from ANs with clustering coefficient $C_{cl}=0.02$.} \label{fig:LocalTopology}
\end{figure}

Thirdly, ANs possess great local topological conditions for cooperation to stabilize. Another two local topological quantities: gradient $g$ and topology potential $U$, introduced in \cite{Yang20092750}, have been shown a significant positive correlation with node's strategy preference. The gradient $g_{i}$ of node $i$ is defined as the average Euclidian distance of degrees with its neighbors $g_{i}=\sqrt{\frac{1}{k_{i}}\sum_{j\in G_{i}}(k_{i}-k_{j})^{2}}$, while the topology potential $U_{i}$ is given by coupling node $i$'s degree and gradient, $U_{i}=k_{i}*g_{i}$. When highly connected hubs tend to be occupied by cooperators stably, defectors can only manage to survive at nodes with low $g$ and $U$. The players on nodes with higher $g$ and $U$ will pay more time to cooperate due to the influences from high-payoff cooperating neighbors \cite{Yang20092750}. In other words, higher $g$ or $U$ indicates more stability of cooperation on these nodes. We have checked cooperation stability of hubs on ANs in Fig.~\ref{AN}. And Fig.~\ref{fig:LocalTopology} shows that ANs possess nodes with high $g$ and $U$, which will be demonstrated as an important factor for the promotion of cooperation on ANs below.

Finally, we find that the special underlying global structure--\emph{degree-skeleton} is another important factor for cooperation to fixate on ANs. The degree-skeleton is a tree spanned by the maximal linked hub with gradual decrease of node's degree and traveling the whole network. It describes the condition that any node except the maximal-degree one has at least one higher-degree neighbor. In networks with a degree-skeleton, the asymmetric learning/replicator dynamics can be fully conducted on the degree-skeleton. With cooperation stabilized on hubs, therefore, it will diffuse effectively from large degree nodes to small degree nodes and eventually diffuse over the whole network.

In order to demonstrate last two points, we employ the Xulvi-Brunet--Sokolov algorithm \cite{Xulvi2004} to relink the original ANs. At each step, two edges with four different vertices are randomly relinked. This relink keeps the degree distribution but gradually destroys other local topological features, including $C_{cl}$, $g$ and $U$. The process is repeated in two ways: \emph{randomly relinking} or \emph{keeping degree-skeleton}, both until an identical given $C_{cl}$ is achieved, which detaches the effect of clustering coefficient. The latter holds a degree-skeleton, which somehow keeps to a certain degree the above advantages: higher $g$ and $U$ as shown in Fig.~\ref{fig:LocalTopology}. For comparison, we rebuild a random degree-skeleton for randomly relinked networks by attaching every local degree-maximum node to any higher-degree node, which belongs to the set of nodes spanned by the maximal linked hub through the pathway decreasing degree \footnote{Notice that we need to check that the attached node still keeps at least one higher-degree neighbor.}. These shortcuts change the global connection totally, but with few influence on all properties of local topological conditions (Fig.~\ref{fig:LocalTopology}). Thus, degree-skeleton rebuilt networks possess a degree-skeleton compared to randomly relinked networks, but lower $g$ and $U$ than degree-skeleton kept networks.

\begin{figure}[t]
  \includegraphics[width=2.5in]{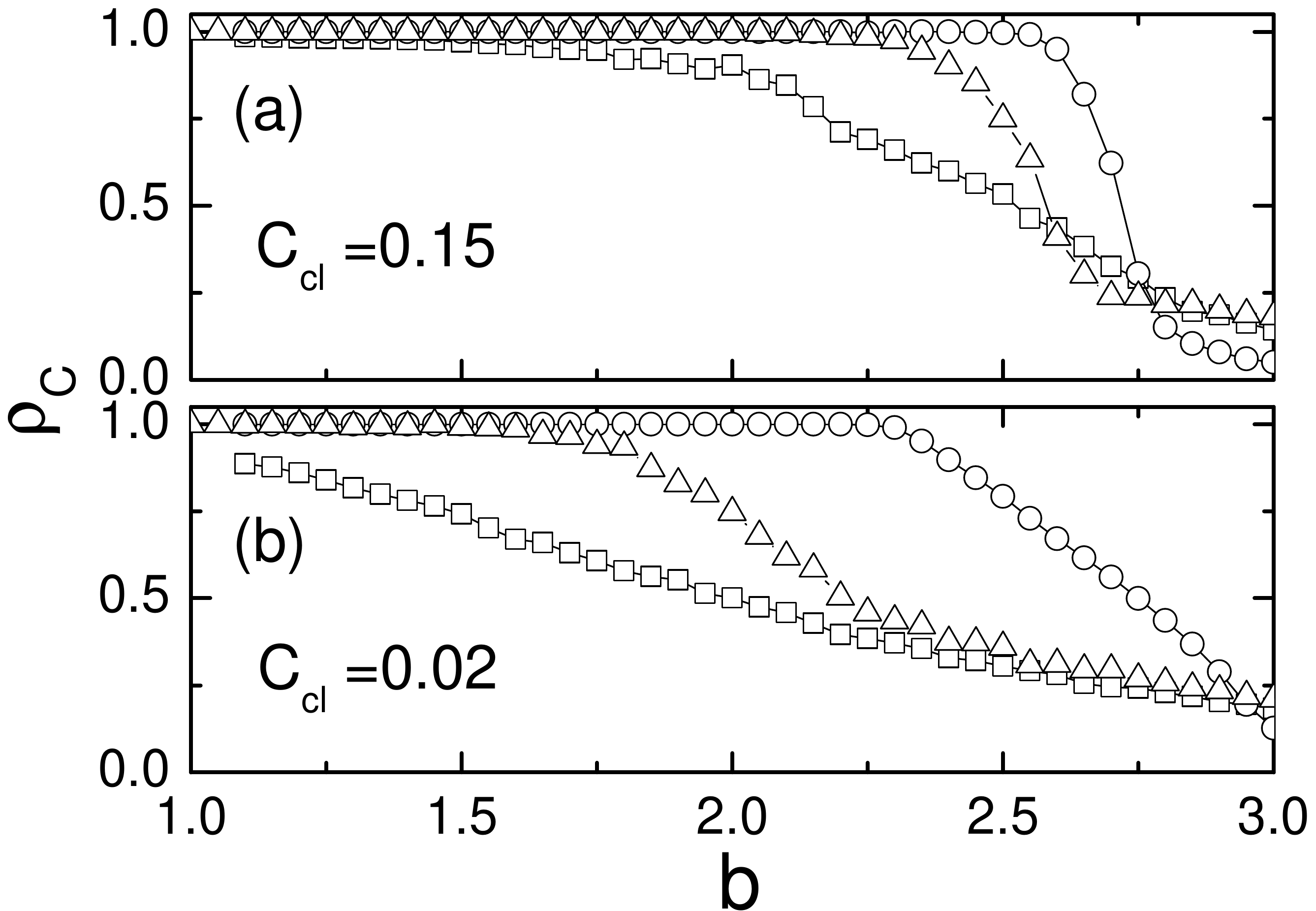}\\
  \caption{Density of cooperators $\rho_{C}$ \emph{vs} temptation $b$ for randomly relinked networks (square), degree-skeleton rebuilt networks (triangle) and degree-skeleton kept networks (circle), which are derived from ANs with various identical $C_{cl}$ and identical degree distribution. }\label{KSvsAR}
\end{figure}

Randomly relinking destroys the degree-hierarchical structure and gives rise to many nodes to be local degree-maxima. The resulting random scale-free networks create some particular conditions where defectors can manage to survive \cite{Gomez2008296}. Such topological property hinders the cooperation to spread and induces the competition between local degree-maxima nodes as the case of dipole model \cite{Floria2009}. At $C_{cl}=0.02$, certainly, the dependency of $\rho_{C}$ on $b$ (Fig.~\ref{KSvsAR}(b), square) can be well understood in terms of the cooperation enhancing properties in the heterogeneously random networks \cite{Devlin2009a}. From this, we are sure that the fixation of cooperation in ANs must not be simply an outcome of having higher hubs. And higher cooperation level at $C_{cl}=0.15$ (Fig.~\ref{KSvsAR}(a)) confirms the above point: the effect of clustering coefficient in cooperation enhancement.

Rebuilding a random degree-skeleton to the above networks enhances cooperation a lot and induces full cooperation at low $b$ (Fig.~\ref{KSvsAR}, triangle), whatever the networks are highly or lowly clustered. The degree-skeleton provides a pathway for cooperation stability to spread and sweeps away defector's surviving space. Specially, for degree-skeleton kept networks, the special degree-skeleton underlaid in ANs promotes cooperation significantly by organizing the advantageous local topology conditions (higher $g$ and $U$) for cooperation stability. The fixation of cooperation on them can be hold to much larger $b$, even though $C_{cl}$ is reduced to $0.02$ (Fig.~\ref{KSvsAR}(b), circle). These results indicate that degree-skeleton improves cooperation condition indeed and the association of degree-skeleton and higher $g$ and $U$ can enhance cooperation further. So we can arrive at the conclusion that the exactly special degree-skeleton underlying ANs provides a strong power for cooperation to fixate \footnote{We have checked that just keeping connections of first 4 hubs doesn't promote cooperation level on randomly relinked network. Thus the function of degree-skeleton on cooperation enhancement doesn't just lie in the connections of hubs.}.

\begin{figure}
  \includegraphics[width=3in]{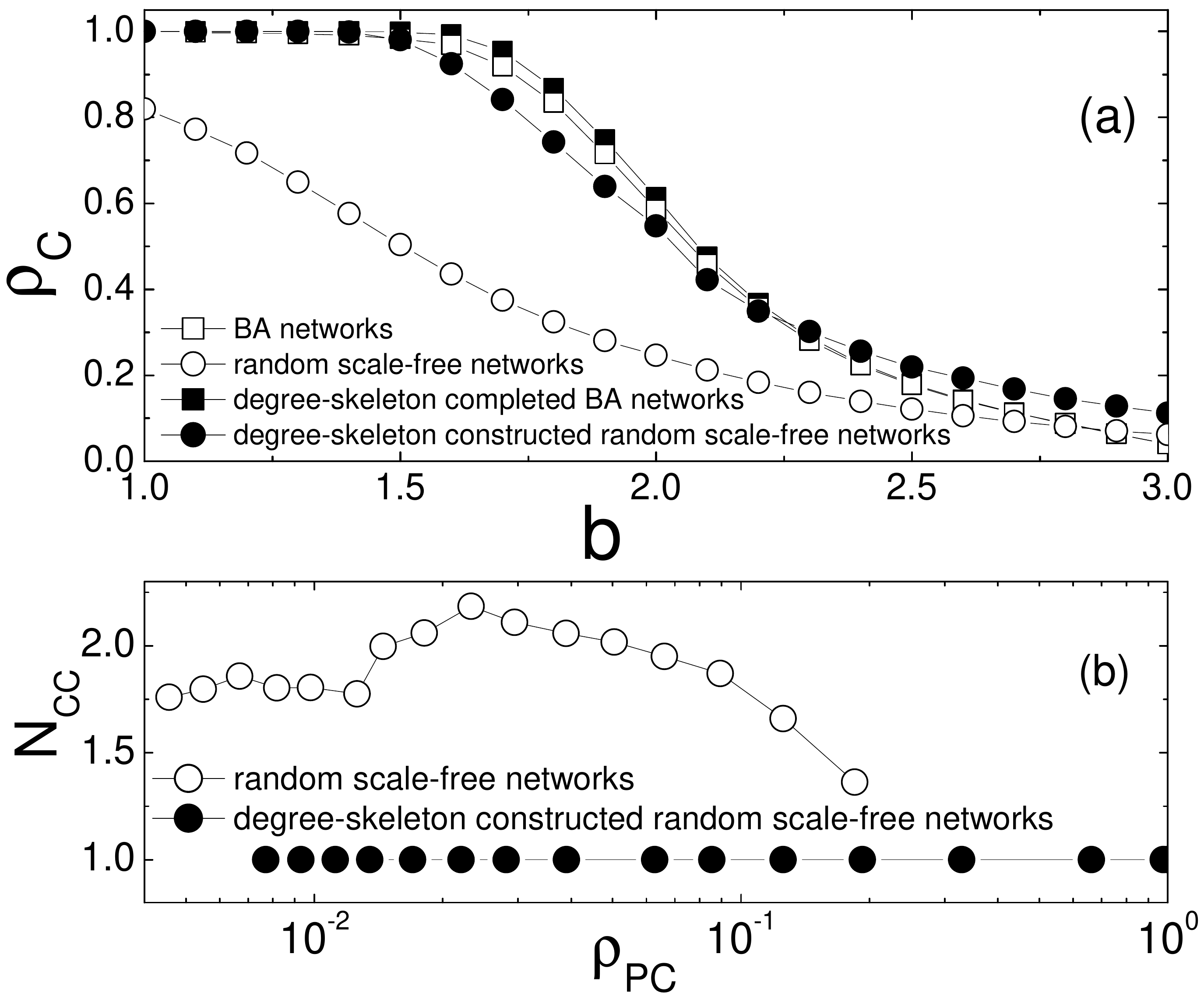}\\
  \caption{(a) Density of cooperators $\rho_{C}$ \emph{vs} temptation $b$ for four networks with or without a full degree-skeleton. (b) Cluster number of pure cooperators \emph{vs} the fraction of pure cooperators $\rho_{PC}$ for random scale-free networks and degree-skeleton constructed random scale-free networks, respectively. The figure clearly shows that by constructing a degree skeleton for random scale-free networks, clusters of pure cooperators can merge together and then promote cooperation significantly.}\label{BAscale-free}
\end{figure}

The deeper insight is that degree-skeleton can naturally lead to the exact one cluster of pure cooperators (the constant cooperators in all rounds after transient time), which was called as the cooperator core in Ref.~\cite{Gomez2007}. Specially in BA networks, the single cluster of pure cooperators \cite{Gomez2007} is also supposed to be induced by the underlying degree-skeleton. Here, we have checked that due to the age-correlation effect, a degree-hierarchical subnetwork spanned by the highest-degree hub indeed covers almost all nodes. There are just $2\sim 5$ local degree-maxima nodes and $10\sim 30$ nodes of $10^4$ nodes excluded outside that subnetwork. Completing a degree-skeleton for BA networks makes almost no difference in the promotion of cooperation (Fig.~\ref{BAscale-free}(a)). In contrast, in random scale-free networks \cite{Molloy1995}, which usually have many local degree-maxima nodes, much lower cooperation level is achieved. However, constructing a degree-skeleton for these networks enhances the cooperation level close to that of BA networks (Fig.~\ref{BAscale-free}(a)) by merging clusters of pure cooperators together (Fig.~\ref{BAscale-free}(b)). With cooperation stability on hubs, thereby, degree-skeleton indeed provides an important pathway for cooperation stability to diffuse on non-clustered scale-free networks. These results imply that degree-skeleton is indeed an essential element for enhancement as well as organization of cooperation on scale-free networks.

In summary, we have found perfect networks--ANs for cooperation to evolve and even fixate, as long as just 3 or 4 cooperators taking up the most-linked hubs. ANs are highly clustered scale-free network with lower degree exponent and much better local topological conditions for cooperation to stabilize. Importantly, our results unveil a special degree-skeleton underlying ANs providing a pathway for cooperation to spread over the whole network. Degree-skeleton is also vital in the mechanism of promoting cooperation on BA networks compared to random scale-free networks, which can also get as high cooperation level as BA networks just by constructing a random degree-skeleton. Due to usually asymmetrical interactions between hubs and other nodes, the heterogeneous networks may induce many remarkable phenomena for various dynamics on them, such as signal transduction and response, synchronization and transport and so on \cite{Huang2006,*Acebron2007,*Gomez2007b,*Gomez2011}. It is worthwhile to investigate the function of degree-skeleton on different kinds of dynamics performed on scale-free networks.

\begin{acknowledgments}
This work is supported by the National Science Foundation of China under Grant No. 30970970.

\end{acknowledgments}
\bibliography{Game}

\end{document}